\documentclass{elsart}

\usepackage{harvard}

\def\astrobj#1{#1}

\def\gaeq{\stackrel{>}{\scriptstyle \sim}}

\begin{document}

\begin{frontmatter}
\title{``Real-Time'' Evolution in Mira Variables}
\author{Patricia A Whitelock}
\address{SAAO,
P O Box 9, Observatory, 7935, South Africa.\thanksref{email}}
\thanks[email]{E-mail: paw@saao.ac.za}

\begin{abstract}
 After a brief review of our current understanding of Miras and their
evolutionary status, three aspects of ``real-time'' evolution in these and
related stars are examined: \begin{itemize}
 \item Chemical changes (O-rich to C-rich) due to third dredge-up
\item Period changes due to the effects of the helium-shell flash
\item The existence of ``fossil'' dust and gas shells. \end{itemize}
Studies of resolved gas and dust shells are highlighted as particularly
interesting. They will enable us to examine the mass-loss histories of many
late-type stars over the last ten thousand years or so.  Such observations
have only recently become technically feasible and they are expected to
provide important new insights into the late stages of stellar evolution.
\end{abstract}

\end{frontmatter}


\section{Introduction}
 Miras are large-amplitude long-period cool variables. Their luminosities
and surface chemical compositions indicate that they are at the very tip of
the Asymptotic Giant Branch (AGB). Their progenitors must have been low- or
intermediate-mass stars and their next evolutionary move will be a rapid
crossing of the Hertzprung--Russell diagram to become white dwarfs, possibly
via a planetary nebula phase. The lifetimes of Miras are not well known, but
for low-mass low-metallicity systems they are probably around $2\times 10^5$
years. The sources of nuclear power for these stars are hydrogen and helium
shells surrounding an inert carbon-oxygen core. There are two processes
which influence stellar evolution towards the end of the AGB: mass loss and
helium-shell flashes. Both are poorly understood as is their interaction.
All Miras are undergoing mass-loss at significant rates although these vary
by three orders of magnitude from one star to another and probably within a
single star at various times.  The crucial thing about mass-loss in this
context is that it removes material from the outer envelope more rapidly
than nuclear processing removes it from the inner envelope. As the size of
the envelope ultimately determines the AGB lifetime, mass loss is an
important evolutionary influence.  Most Miras undergo helium-shell flashes
following the build up of helium from the hydrogen burning shell. This is
most conclusively demonstrated by the presence of the short-lived isotope of
technetium, \nuc{99}{Tc}, in their spectra. Technetium must have been brought to the
surface by dredge-up following a helium-shell flash. A few short period
Miras have no clear signs of chemical anomalies, and may or may not
be undergoing helium-shell flashes. If they are, then either dredge-up has
not accompanied the flash or the time interval between flashes and dredge-up
is sufficiently long that much of the technetium has decayed.

There are a number of important theoretical papers dealing with AGB
evolution, e.g. \citeasnoun{IR83}, \citeasnoun{Lattanzio}, \citeasnoun{BS88},
\citeasnoun{VW93}. \citeasnoun{VW93} discuss, for a low mass star, the
changes in mass-loss rate and surface luminosity during a sequence of shell
flashes. They find that heavy mass-loss ($\dot{M} > 10^{-6}\ M_{\odot}\,{\rm
yr^{-1}}$) occurs only during the last two or three flash cycles prior to
AGB termination. Most models discussed to date give a typical time between
late pulses in a low mass star of about $1-3 \times 10^5$ years, e.g.
\citeasnoun{BS88}, \citeasnoun{VW93}. The interval depends on composition and is shorter
for higher mass stars. If we are correct in assuming that Mira variables
represent the high luminosity and high mass-loss phase of AGB evolution,
then their lifetimes ($\sim 2 \times 10^5$ years) tell us that they cannot
last more than a very few helium-flash cycles. Exactly how many cycles
depends on what fraction of the inter-pulse period they actually spend as
Miras. The models indicate significant changes, $\sim$1 mag, in the
bolometric magnitude during the inter-pulse phase (even larger changes occur
during the very brief helium-shell flash itself). The narrow period luminosity
relation which is observed for low mass stars \cite{Feast89,Whitelock94} is
not consistent with stars remaining Miras throughout these luminosity
changes. We are therefore forced to the conclusion that stars move in and
out of the Mira instability strip during the last few helium-shell flashes.

The remainder of this paper, following a brief caveat about binary stars, is
concerned with observational evidence for changes which may be a direct
consequence of evolution near the top of the AGB.  Specifically we examine
evidence for chemical changes produced by the third dredge-up which follows
certain flashes, for period changes that accompany the extreme luminosity
changes, and for ``fossil'' dust shells which become isolated from the star
during the flash cycle.

\section{Confusion with Binary Evolution}
 It is worth noting that the existence of binaries is a potential source of
confusion in our understanding of luminous AGB stars. Binary mass-transfer
and common envelope evolution can give rise to effects very similar to those
anticipated from helium-shell flashes in single stars, e.g. enhanced
mass-loss rates. There are many possibilities of confusion between the
effects seen in binary interaction and in single star AGB or post-AGB
evolution. Relatively wide binaries, those with orbital periods of a few
years, will interact for the first time when one of the stars reaches the
extreme size of a Mira variable (diameter $\gaeq 1\, $AU).

\section{Chemical Changes}
 A variety of chemical changes are expected as a result of the third
dredge-up thought to follow helium-shell flashes under certain
circumstances, e.g. \citeasnoun{IR83}. It is extremely difficult to look for
minor chemical changes in Miras because large changes in excitation
conditions around the pulsation cycle are quite normal. It is not unusual
for the colour temperature to change by 1000\,K between maximum and minimum
light. Furthermore, changes do not repeat exactly from cycle to cycle, so
even repeated observations at the same phase can show significant
differences. However, there are two possible cases of changes involving the
C/O ratio going from less than one to greater than one, for \astrobj{TT~Cen}
and \astrobj{BH~Cru}. In other words, we see a star apparently change from
oxygen rich to carbon rich in the way that we might expect if the products
of third dredge-up were brought to the surface in real time. The evidence
for \astrobj{TT~Cen} is rather limited and the early oxygen-rich
classification is based on objective prism spectra. It must therefore be
regarded as somewhat uncertain. The case for \astrobj{BH~Cru} seems to be
stronger. This star had been classified as SC, with strong ZrO and no $\rm
C_2$, in 1965-73 by \citeasnoun{FC71} and by \citeasnoun{KB80}. In 1980
\citeasnoun{Evans85} found strong $\rm C_2$ bands. He has since been
following the star around its pulsation cycle and finds it to be a stable CS
star (Lloyd Evans, private communication). Interestingly, there is also some
suggestion from the RASNZ that the period of \astrobj{BH~Cru} changed from
421 day in the 1970s to over 500 day in the 1990s \cite{Bateson88,Walker95}.
The periods of Miras tend to be rather unstable and it is not yet clear if
this variation can be attributed to evolutionary changes (see below).

\section{Period Changes}
 Period changes are an obvious evolutionary effect to look for in Miras, and
for this the databases of the various amateur variable monitoring groups are
invaluable . Looking for period changes in these stars is not as simple as
one might suppose. Both the periods and the amplitudes of many of these
stars change in a variety of ways, sometimes drifting slowly sometimes
making discontinuous large changes \cite{KL}. They do this on a variety of
time-scales. Of particular interest are the rare stars which show clear
period changes and might therefore plausibly be undergoing structural
changes associated with helium-shell flashing. Two of these have been known
for a long while: \astrobj{R~Aql} and \astrobj{R~Hya} were noted as having
changing periods in the 1930s and the trends have continued. They change at
about 1.0 and 1.3 day per year, respectively, and have been doing so for over
a century. \citeasnoun{WZ81} tentatively added \astrobj{W~Dra} to this list
and interpreted the changes in terms of the effects of helium-shell flashes.
\citeasnoun{Percy} have suggested that three more stars, \astrobj{S~Ori},
\astrobj{Z~Tau} and \astrobj{Z~Aur} may also have changing periods. Recently
two groups, \citeasnoun{GS} and \citeasnoun{MF} have suggested that
\astrobj{T~UMi} be added to the list. Its period was roughly constant at
about 315 day from 1905 until 1979. Since then it has been decreasing,
reaching 274 day in 1994. It is too soon to say that this star is in the
same category as \astrobj{R~Aql} and \astrobj{R~Hya}; a longer time base is
essential for reasonable certainty about evolutionary change.

\citeasnoun{WZ81} discuss how the slow changes observed in \astrobj{R~Aql},
\astrobj{R~Hya} and \astrobj{W~Dra} might fit with the luminosity changes
anticipated during the inter-pulse phase of the flash cycle. In contrast
\citeasnoun{MF} suggest that \astrobj{T~UMi} is in the early helium-shell
flash phase where the period is expected to change most rapidly as the
luminosity plummets.  Although, as mentioned above, it is too soon to be
sure of what is occuring in \astrobj{T~UMi} it is a sufficiently interesting
star that intensive observations of it would be worthwhile. If it is indeed
heading towards the luminosity minimum at the start of a helium-shell flash
then we should anticipate many changes in the near future, including perhaps
a reduction of the pulsation amplitude as it drops out of the Mira
instability strip.  The databases now contain observations for large numbers
of Miras over a century or more. It would be interesting to see a detailed
statistical analysis of these data for many stars. It might then be possible
to put some rather definite limits on the period changes of Miras in the
inter-pulse phase. A preliminary analysis of this sort has been reported by
\citeasnoun{Percy96}. It is worth noting that the time interval covered by a
helium-shell flash is so short that we would not expect to find more than
one or two stars undergoing a flash among a sample of 1000 Miras.

\section{``Fossil'' Shells}
 One of the most interesting recent developments in this field has been the
discovery and measurement of the extended circumstellar environments of
Miras and related stars. In terms of our present discussion these essentially
present a ``fossil record" of the mass-loss from the star over the preceding
few thousand years. This is a subject in its infancy which could profoundly
change our understanding of AGB evolution and mass loss. There is
observational evidence for ``fossil'' shells from extended CO radio
emission, dust shells  resolved at infrared wavelengths, resonance
scattering \cite{Gustafsson} and even optical nebulosity
\cite{Whitelock}. Of particular interest is the discovery of detached dust
or gas shells which indicate a severe decrease in the mass-loss rate in
recent times. This is discussed below with emphasis on the evidence from
high resolution CO observations for detached gas shells and from the
infrared for detached dust shells. The above discussion was limited to Mira
variables, but in the case of detached dust-shells it is important to 
examine also the semi-regular (SR) variables, on the assumption that those with
detached shells were Miras in the recent past. It is, however, likely that
they had somewhat more massive and/or younger progenitors than had most of
the Miras discussed above. Note that \astrobj{CW~Leo} is a Mira with a shell
which is resolved at many wavelengths. It is not clear if this star fits the
general pattern and it is at least possible that it is a binary system.

Large numbers of carbon variables have been examined for CO emission by
Olofsson and coworkers, e.g. \citeasnoun{Olof}, and a small proportion, about
6\%, have been shown to have extended shells. In general the distribution of
CO is essentially spherical although often rather clumpy. The shell around
the SR variable \astrobj{TT~Cyg}, one of the most extended sources,
appears to be expanding at about 12 $\rm km\, s^{-1}$ although material
currently being lost from the star itself is moving at less than half that
velocity. If \astrobj{TT~Cyg} is at its Hipparcos distance of 510 pc then
the shell is about 7 thousand years old. The mass-loss rate when the shell
was ejected must have been around $\rm 10^{-5}\ M_{\odot}\, {\rm yr^{-1}}$,
although current values are two orders of magnitude smaller.

 Some red giants show 60 and $100\,\mu$m excess in their IRAS flux
distributions. This is more common among carbon-rich than among oxygen-rich
stars. It has been interpreted as the result of a previous high mass-loss
phase among stars which currently have relatively low mass-loss. The
situation is complicated by the fact that some of the 60 and $100\, \mu$m
excesses are not intrinsic but the result of contamination of normal
stellar emission by infrared cirrus. However, there have now been enough
detailed studies to be sure that there are many real detached dust shells. 
The SR variable \astrobj{U~Hya} is an example. A maximum entropy analysis of
the IRAS $60\, \mu$m maps \cite{Waters} reveals a dust ring about 3.5
arcmin in diameter at a temperature of about 50\,K. It is presumed to be the
inner part of the dust shell because the outer part will be too cool to radiate
at $60\, \mu$m. The authors estimate that the shell is about $12\,000$ years
old and was formed when the mass-loss rate was $\dot{M} \sim 5 \times
10^{-6} \ M_{\odot}\, {\rm yr^{-1}}$, which is typical of a Mira and about
25 times higher than the present value.

\citeasnoun{Izumiura} used the same maximum entropy technique to examine the
IRAS data for \astrobj{U~Ant}, an L-type variable. They found a more complex
structure and suggest that we are seeing a detached dust ring in addition to
the obvious resolved material surrounding the central star.  Izumiura (1997
private communication) has since confirmed the existence of the outer ring
using ISO images at $90\, \mu$m. For \astrobj{U~Ant}, \citeasnoun{Izumiura}
suggest the time interval between the two high mass-loss phases is only
about $3\,000$ years, compared to more than $12\,000$ years for
\astrobj{U~Hya}. If this is the interval between helium-shell flashes then
it would imply a star with a relatively high initial mass of 4 to 5
$M_{\odot}$.

\ack I am grateful to Michael Feast for many helpful discussions and to H.
Izumiura and T. Lloyd Evans for discussing their work in advance of
publication.



\end{document}